\documentclass[twocolumn,aps,description]{revtex4}

\usepackage[colorlinks,linkcolor=blue,citecolor=blue,urlcolor=blue]{hyperref}

\usepackage{graphicx}
\usepackage{epstopdf}
\usepackage{amsmath}
\usepackage{multirow}

\begin{document}

\title{Magnetic surface on nonmagnetic bulk of electride Hf$_2$S}

\author{Jian-Feng Zhang$^{1,2}$}
\author{Duo Xu$^1$}
\author{Xiao-Le Qiu$^1$}
\author{Ning-Ning Zhao$^1$}
\author{Zhong-Yi Lu$^1$}\email{zlu@ruc.edu.cn}
\author{Kai Liu$^1$}\email{kliu@ruc.edu.cn}

\affiliation{$^1$Department of Physics and Beijing Key Laboratory of Opto-electronic Functional Materials $\&$ Micro-nano Devices, Renmin University of China, Beijing 100872, China}
\affiliation{$^2$Institute of Physics, Chinese Academy of Sciences, Beijing 100190, China}

\date{\today}

\begin{abstract}
Recent experiment reported the self-passivated electride Hf$_2$S with excellent stability and continuous electrocatalytic ability [S. H. Kang \textit{et al.}, Sci. Adv. \textbf{6}, eaba7416 (2020)]. Starting from its 2H-type layered structure, we have studied the electronic, magnetic, and transport properties of the electride Hf$_2$S in the monolayer and multilayer forms by combining first-principles electronic structure calculations and Kubo formula approach. Our calculations indicate that these thin films of Hf$_2$S electride are both dynamically and thermodynamically stable. Astonishingly, the calculations further show that the outmost Hf atoms and the surface electron gas of the Hf$_2$S multilayers are spin polarized, while the inner Hf atoms and the electron gas in the interlayer regions remain nonmagnetic. Due to the magnetic surface, the multilayer Hf$_2$S exhibits many unusual transport properties such as the surface anomalous Hall effect and the electric-field-induced layer Hall effect. Our theoretical predictions on Hf$_2$S call for future experimental verification.
\end{abstract}


\maketitle

Electrides, identified by the anionic electrons locating at the vacant crystallographic sites and being unbounded to any nuclei in the lattice, have attracted widespread attention recently~\cite{review1,review2}. The diverse distributions of interstitial electrons in the electrides endow them with particular properties, including high electronic concentration and conductivity, high density active sites, low work function, etc~\cite{review1,review2,concen,workfunc}. These outstanding properties make electrides serve as excellent candidates for the applications in catalysts, rechargeable batteries, electron emitters, light-emitting diodes, and so on~\cite{review1,review2,catalysts,emission,emitter}. Besides the practical applications, many interesting quantum phenomena, such as superconductivity~\cite{supercon}, magnetism~\cite{Gd2C1}, and nontrivial topological property~\cite{wengsb}, have also been discovered in the electrides. And the role of the interstitial electrons playing in the those properties of electrides has been revealed. Nevertheless, the research on the electrides is still in its infancy and there are still plenty of properties remaining unexplored.


According to the space distribution property of its interstitial electron gas, electride can also perform dimensionality characters (0$\sim$3D)~\cite{maprx}. Specially, the 2D electrides with layered structures have attracted intensive attention due to their unique physical properties, feasible exfoliation down to atomically thin films, as well as potential integrations with other 2D materials. For instance, versatile topological band features were found in the 2D electrides such as Y$_2$C, Sc$_2$C, etc~\cite{Y2C,topoprx}. The nonmagnetic 2D electride Ca$_2$N~\cite{Ca2N1} may become magnetic via the hydrogenation~\cite{Ca2N}. The ferromagnetic 2D electride Gd$_2$C~\cite{Gd2C} owns high Curie temperature~\cite{Gd2C1} and multiple pairs of Wely points~\cite{Gd2C}. Recently, it was reported that a self-passivated 2D electride namely dihafnium sulfide, written as [Hf$_2$S]$^{2+}$$\cdot$2$e^-$, exhibits a strong oxidation resistance in water and acid solutions and can persist the electrocatalytic hydrogen evolution reaction~\cite{Hf2S}. Since Hf$_2$S owns a 2H-type layered structure, whether its thin-film forms are stable and whether it hosts interesting quantum phenomena as the reduction of dimensionality wait for further indepth studies.

In this work, based on first-principles electronic structure calculations and Kubo formula approach, we have studied the electronic, magnetic, and transport properties of the multilayers of Hf$_2$S electride. Our calculations indicate that distinct from the nonmagnetic bulk phase, multilayer Hf$_2$S can hold a magnetic electron gas at the surface along with the nonmagnetic electron gas in the inner interlayer regions. We then analyzed the origin of the surface magnetism, and further studied the electric-field-induced layer Hall effect and the abnormal Hall effect of multilayer Hf$_2$S.

The structural, electronic, and magnetic properties of $n$-layer ($n=$ integer) Hf$_2$S were investigated with the spin-polarized density functional theory (DFT)~\cite{dft1,dft2} calculations as implemented in the Quantum ESPRESSO (QE)~\cite{pwscf} and VASP~\cite{vasp1,vasp2} packages. The generalized gradient approximation of Perdew-Burke-Ernzerhof~\cite{pbe} (PBE) type was adopted for the exchange-correlation functional.
The 2D anomalous Hall conductivity ($\sigma_{xy}^{\text {2D}}$) was calculated with the Kubo-formula approach in the linear response scheme~\cite{AHC} via the Berry curvature by using the WannierTools package~\cite{wt}, which is interfaced to the Wannier90 software~\cite{mlwf}.
More computational details are described in the Supplemental Information (SI)~\cite{sm}.

\begin{figure*}[t]
\includegraphics[angle=0,scale=0.6]{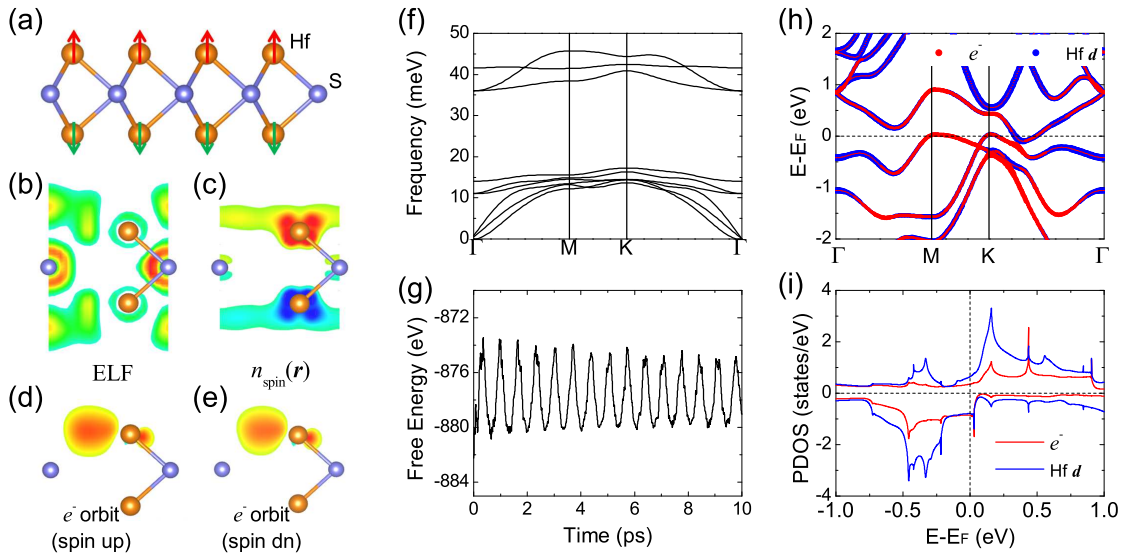}
\caption{(Color online) (a) Side view of monolayer Hf$_2$S. The orange (purple) balls represent the Hf (S) atoms. The red (green) arrows label the up (down) spins. Spatial distributions of (b) the electron localization function (ELF) and (c) the spin density $n_{\rm spin}({\bf r})=n_{\uparrow}({\bf r})-n_{\downarrow}({\bf r})$. The red and blue regions in panel (c) indicate the signs of $n_{\rm spin}({\bf r})$. The orbital distributions derived from the maximally localized Wannier functions (MLWF) for the top-surface electron gas $e^-$ in the (d) spin-up and (e) spin-down channels. (f) Phonon dispersion. (g) Time evolution of free energy at 300 K from a molecular dynamic simulation. (h) Electronic band structure. The sizes of red and blue dots on the bands represent the weights of the $e^-$ and Hf-$d$ MLWF orbitals. (i) Partial density of states (PDOS) for the $e^-$ (red lines) and Hf $d$ (blue lines) MLWF orbitals in the top surface. The upper and lower parts correspond to the spin-up and spin-down channels, respectively.}
\label{fig1}
\end{figure*}

We firstly studied Hf$_2$S in the monolayer limit. From the side view of monolayer Hf$_2$S [Fig. 1(a)], we can see that the S atomic plane is sandwiched by two Hf atomic planes with the mirror symmetry. In each atomic plane, S or Hf atoms form a triangular lattice. Considering the partially filled 5$d$ orbitals of Hf atoms, we performed the spin-polarized DFT calculations to check the possible magnetism on the Hf lattice. Several typical spin configurations have been considered (see Fig. S1 in the SI). We find that the ground state of monolayer Hf$_2$S is an antiferromagnetic state, in which the magnetic moments of the Hf atoms in the same Hf plane are parallel while those between the upper and lower Hf planes are antiparallel[Fig. 1(a)]. In order to investigate the stability of monolayer Hf$_2$S in this AFM ground state, we further calculated the phonon dispersion. As displayed in Fig. 1(f), there is no imaginary phonon mode across the whole Brillouin zone, indicating its dynamical stability. Figure 1(g) shows the time evolution of the free energy of monolayer Hf$_2$S at 300 K from an \textit{ab initio} molecular dynamics simulation, in which the system enters thermodynamical equilibrium after 6 ps. According to a snapshot at 10 ps, we learn that the structure remains intact without the bond breaking and stays in thermodynamical equilibrium, suggesting that monolayer Hf$_2$S is also thermodynamically stable at room temperature.

We then explore the electronic and magnetic properties of monolayer Hf$_2$S in the AFM ground state. As shown by the electron localization function (ELF)~\cite{ELF} in Fig. 1(b), there are electron gases floating on both surfaces of monolayer Hf$_2$S. These electron gases exhibit the AFM spin polarization as indicated by the real-space spin densities [$n_{\rm spin}({\bf r})=n_{\uparrow}({\bf r})-n_{\downarrow}({\bf r})$] in Fig. 1(c), which are different from the nonmagnetic electron gas in the interlayer regions of bulk phase~\cite{Hf2S,Zhang21}. With the maximally localized Wannier function (MLWF) method~\cite{mlwf}, we can describe the electron gas on the surface of monolayer Hf$_2$S by a Wannier orbital ($e^-$), whose spatial distributions in the spin-up and spin-down channels are demonstrated in Figs. 1(d) and 1(e), respectively. The electronic band structure calculated with the MLWF is displayed in Fig. 1(h), which matches well with the DFT result in SI~\cite{sm}. There are several bands crossing the Fermi level, indicating a metallic behavior. The electronic states around the Fermi level mainly originate from the electron gas ($e^-$) and the Hf $d$ orbitals, whose weights scale with the dot sizes on the bands. The corresponding partial density of states (PDOS) are shown in Fig. 1(i), which indicates that the spin polarization of the electron gas ($e^-$) is parallel to that of its neighbor Hf atoms.

\begin{figure}[t]
\includegraphics[angle=0,scale=0.335]{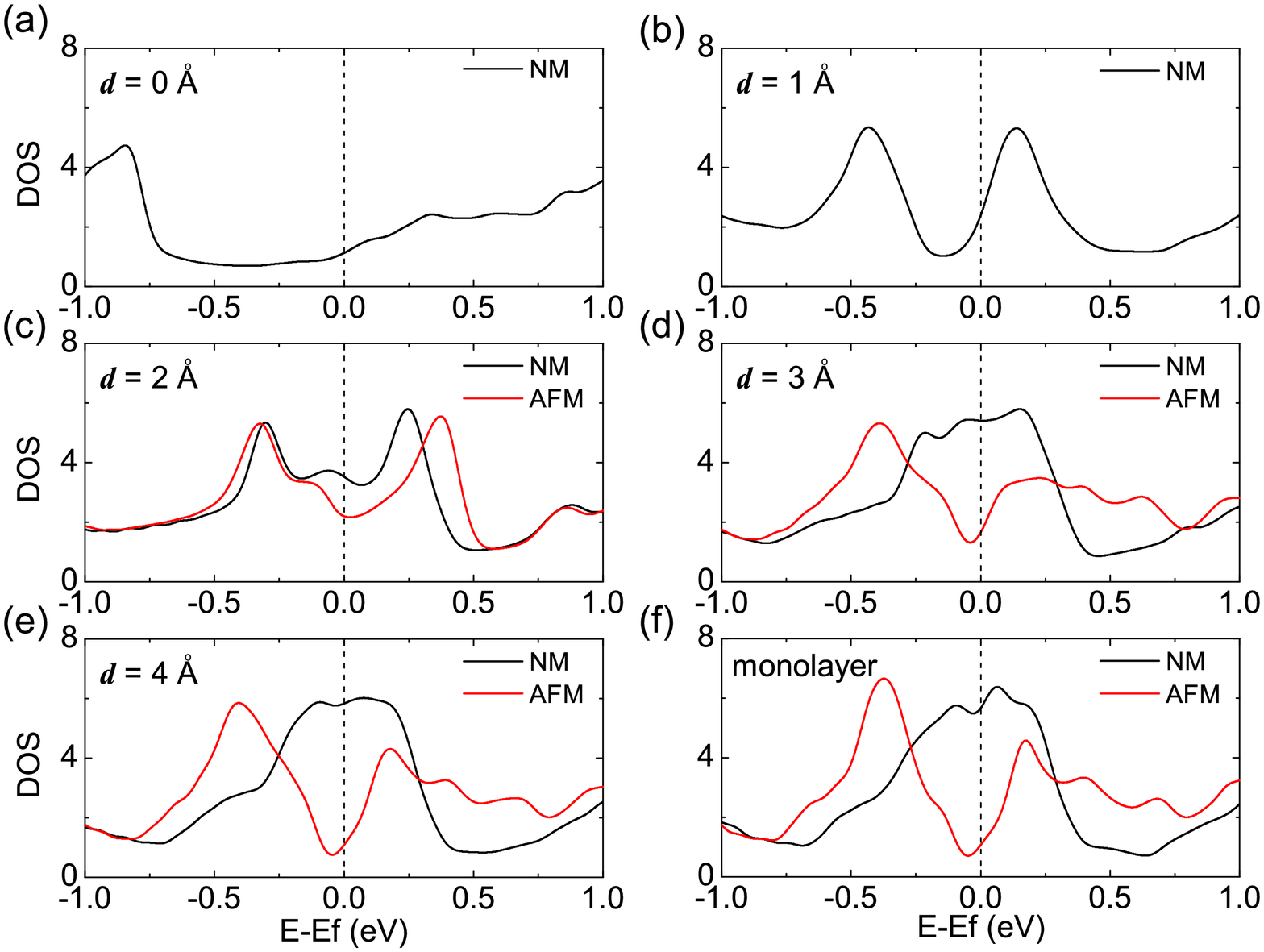}
\caption{(Color online) The evolution of electronic density of states (DOS) (in units of states/eV/f.u.) of Hf$_2$S in the NM (black line) and AFM (red line) states as a function of the interlayer distance $d$. Here (a) and (f) represent the bulk and monolayer cases, respectively. The AFM state emerges from $d$ = 2 \AA~in panel (c).}
\label{fig2}
\end{figure}

To explore the origin of the interesting surface magnetism of monolayer Hf$_2$S, we simulated the exfoliating process from the bulk material to the monolayer limit step by step. Figure 2 exhibits the evolution of electronic DOS with the interlayer distance $d$. In the nonmagnetic bulk phase [Fig. 2(a)], there is no obvious peak around the Fermi level. Once the interlayer distance is enlarged by 1 \AA~($d = 1$ \AA), the DOS shows dramatic changes with two large peaks below and above the Fermi level [Fig. 2(b)]. When the interlayer distance increases to $d=2$ \AA~[Fig. 2(c)], the nonmagnetic DOS (black line) produces a peak at the Fermi level, and the AFM state becomes the ground state and reduces the intensity of DOS at $E_\text F$ (red line). With further increasing $d$ (3 or 4 \AA) [Figs. 2(d) and 2(e)], the peak of nonmagnetic DOS (black line) around the Fermi level is more prominent, which then splits into the upper and lower peaks in the AFM state (red line), forming a V-shaped DOS around the Fermi level. A natural explanation is that the cleavage of the layered structure weakens the original interlayer orbital hybridization, makes the surface electron gas of monolayer Hf$_2$S more localized, and then results in the peak of DOS around the Fermi level.
According to the Stoner criterion, a large DOS around the E$_\text F$ can drive the system into magetic phase, forming FM exchange interaction in same-layered (upper or lower) Hf atoms by itinerant surface electron gas and AFM super-exchange interaction among different Hf atomic layers by middle S atoms. 
When the interlayer distance $d$ becomes large enough, the DOS converges to the monolayer limit as shown in Fig. 2(f).

\begin{figure}[t]
\includegraphics[angle=0,scale=0.41]{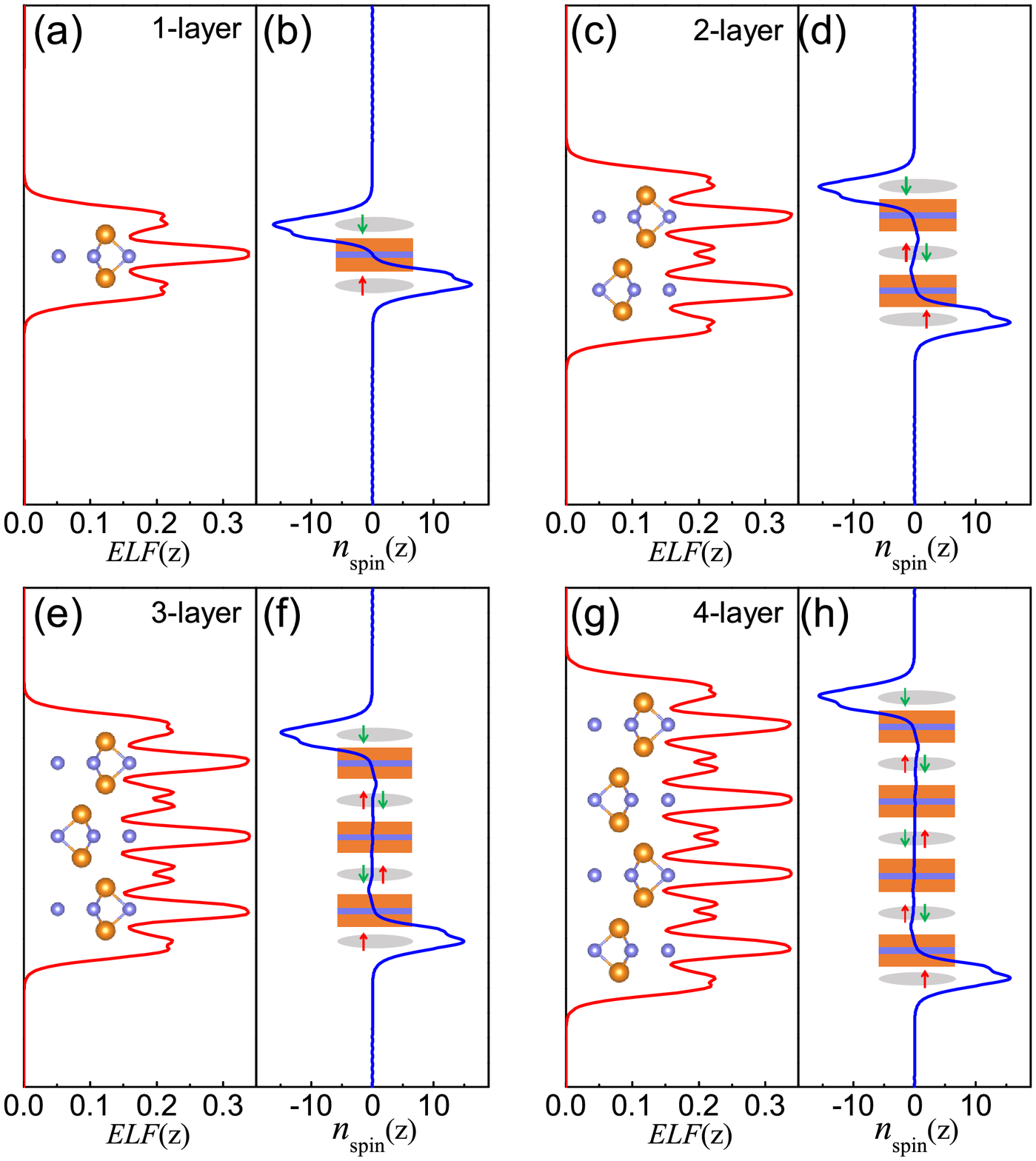}
\caption{(Color online) The calculated electron localization functions (ELF) and spin densities $n_{\rm spin}(\bf r)$ of different Hf$_2$S multilayers: (a-b) 1-layer, (c-d) 2-layer, (e-f) 3-layer, and (g-h) 4-layer. The $x$ and $y$ components were integrated for simple exhibition. The insets in ELF panels display the corresponding atomic structures, while those in $n_{\rm spin}$ panels show the schematic spin distributions.}
\label{fig3}
\end{figure}

Considering that Hf$_2$S is nonmagnetic in the bulk form but AFM in the monolayer phase [Fig. 1(a)], we are curious about the evolution of magnetic properties with the layer thickness. According to the 2H-type stacking of bulk Hf$_2$S, we obtained the atomic structures of 2-layer, 3-layer, and 4-layer Hf$_2$S. The distributions of electron localization functions (ELF) and spin densities $n_{\rm spin}(\bf r)$ in the monolayer and multilayer forms are demonstrated in Fig. 3. From the ELF plots along the $z$ direction, we can see that there are electrons localized on the S planes. Meanwhile, the positions of localized electrons around the Hf atoms slightly deviate from the Hf planes, which results from the accumulation of electron gases in the interstitial regions. Astonishingly, we find that the electron gases locating in the interlayer regions of the Hf$_2$S multilayers are not spin-polarized, while the electron gases at the upper and bottom surfaces still hold their original spin polarizations as in the monolayer form [Figs. 3(d), 3(f), and 3(h)]. The interlayer electron gases in multilayer Hf$_2$S can be viewed as the one obtained by superposing the surface electron gases of two neighboring Hf$_2$S monolayers with the opposite spin polarizations. When the interlayer distance $d$ is small enough, the electron gases with opposite spins from different Hf$_2$S layers coalesce and form a nonmagnetic interstitial one, which is in accordance with its nonmagnetic bulk counterpart. The magnetic electron gas on the surface of the nonmagnetic bulk is an unexpected finding for the electride Hf$_2$S. In real materials, the cleaved Hf$_2$S surface often exhibits the amorphous structure~\cite{Hf2S}. To observe such magnetic surface electron gas, one should obtain a flat and clean surface of Hf$_2$S by using the molecular beam epitaxy (MBE) growth technique~\cite{MBE}.

As the spin-polarized charge only exists on the surfaces (Fig. 3), Hf$_2$S may show novel transport behaviors that resemble but differ from previous Hall effects. In the conventional anormalous Hall effect, an intrinsic ferromagnetic bulk state generates a non-zero anormalous Hall conductivity (AHC) in a zero magnetic field. Here for the Hf$_2$S, the AHC could only exist on the surface since its inner part is nonmagnetic. However, the upper and bottom surfaces of multilayer Hf$_2$S own opposite spin polarizations and there is no net magnetic moment. Hence, a non-zero AHC can just exist on a single surface of multilayer Hf$_2$S. This situation is very similar to the layer-related AHC in MnBi$_2$Te$_4$, so called layer Hall effect (LHE)~\cite{LHE}. In MnBi$_2$Te$_4$, an external electronic field along the $z$ direction (${\bf E}_z$) can break its $PT$ symmetry ($P$: space inversion; $T$: Time reversal) and induce a non-zero 2D AHC ($\sigma_{xy}^{\rm 2D}$)~\cite{LHE}. Similarly, an external ${\bf E}_z$ can also induce the surface-related AHC in multilayer Hf$_2$S. Via the Kubo-formula approach in the linear response scheme~\cite{AHC}, we calculated the ${\bf E}_z$ induced $\sigma_{xy}^{\rm 2D}$ in the $n$-layer Hf$_2$S ($n=1$, 2, 3, and 4). The results have been exhibited in Fig. 4, where the red/blue lines represent those in the opposite ${\bf E}_z$ directions. As can be seen, the calculated $\sigma_{xy}^{\rm 2D}$ of multilayer Hf$_2$S is proportional to the ${\bf E}_z$ strength and it can reach 0.2 $e^2/h$ around the Fermi level at ${\bf E}_z$ = 0.005 a.u. (1 a.u. = 51.422 V/nm). Meanwhile, the reversal of ${\bf E}_z$ can also flip the sign of $\sigma_{xy}^{\rm 2D}$. These features are quite similar to those in MnBi$_2$Te$_4$~\cite{LHE}.

\begin{figure}[!t]
\includegraphics[angle=0,scale=0.32]{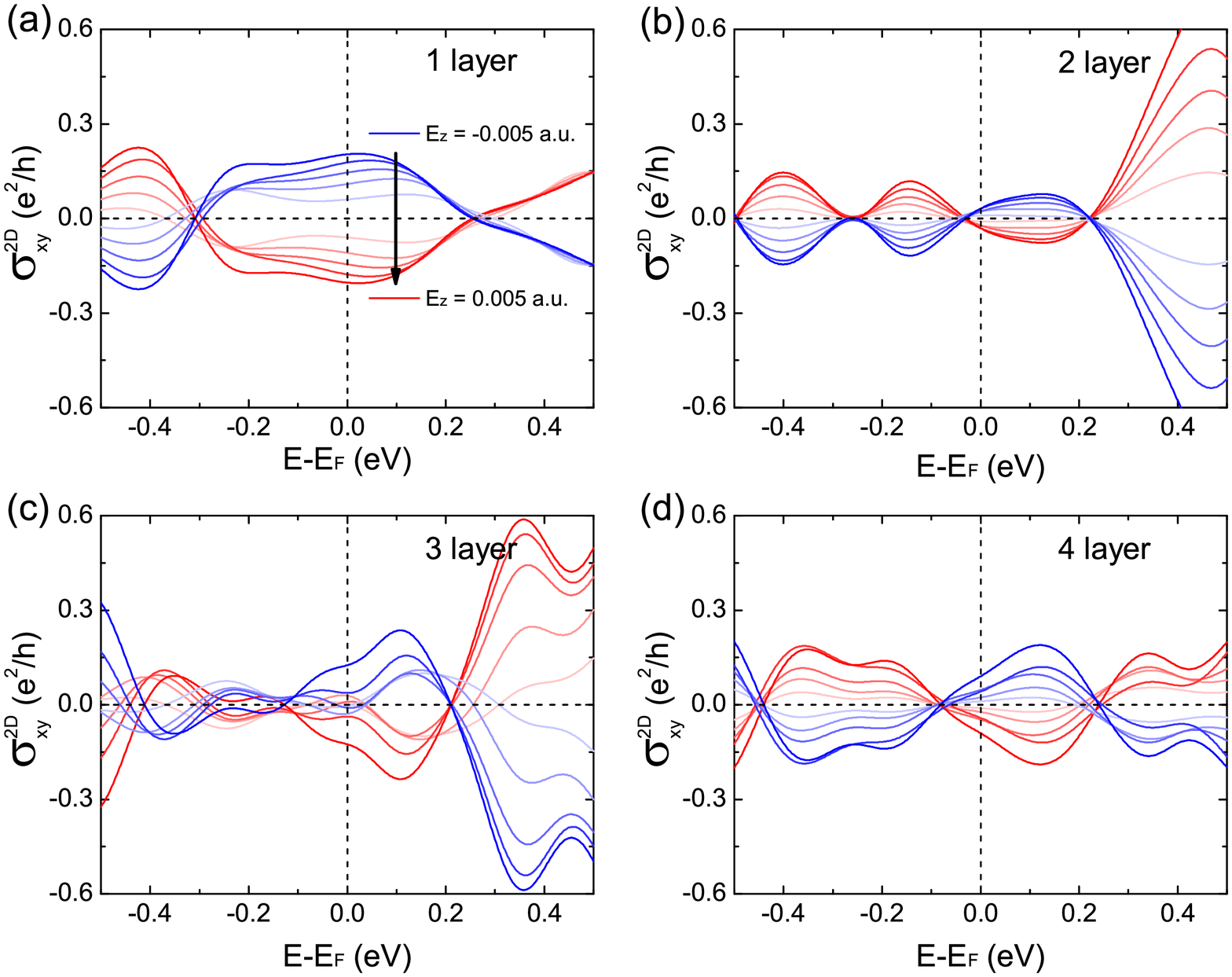}
\caption{(Color online) Calculated 2D anomalous Hall conductivities $\sigma_{xy}^{\text {2D}}$ of (a) 1-layer, (b) 2-layer, (c) 3-layer, and (d) 4-layer Hf$_2$S in the electric field ${\bf E}_z$. The unit of $\sigma_{xy}^{\text {2D}}$ is renormalized to $e^2/h$. The blue lines are symmetrized from the red lines to show the effect of opposite electric field. The color shade is proportional to the field strength.}
\label{fig2}
\end{figure}

In addition to above few layer cases, the surface-related AHC can be manifested more clearly in a thick slab. We constructed a 10-layer thick slab and confirmed that the AFM interaction between its two surfaces is very weak (Fig. S2 in SI)~\cite{sm}. By setting the spins of the two surfaces in the same polarization, we can approximatively study the AHC on a single surface of Hf$_2$S. Figure 5(a) shows the calculated band structure of the ferromagnetic 10-layer Hf$_2$S. The size of color dots represents the orbital weight of Hf atoms in the spin-up (red) and spin-down (blue) channels, respectively, and the nonmagnetic bulk states are also underlaid as the grey background. The spin-polarized surface states and the nonmagnetic bulk states can be easily distinguished, which also indicates that the spin-polarized electronic states only exist on the surface. The calculated 2D AHC on the single surface is shown in Fig. 5(b). The $\sigma_{xy}^{\rm 2D}$ at the Fermi level is about 0.6 $e^2/h$, close to that of the previously reported 2D ferromagnets, such as Fe$_3$GeTe$_2$ (0.4 $\sim$ 1.5 $e^2/h$)~\cite{FGTexp,FGT}.

\begin{figure}[thb]
\includegraphics[angle=0,scale=0.33]{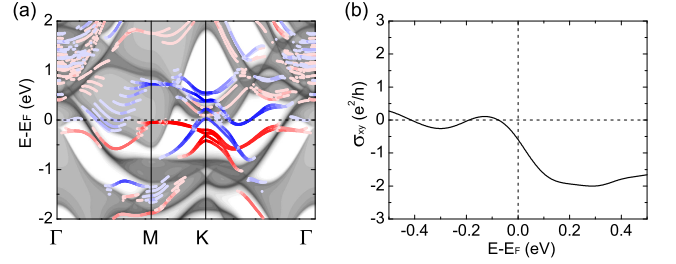}
\caption{(Color online) (a) The ferromagnetic surface state on a 10-layer slab of Hf$_2$S. The color shade of the red and blue dots represent the weight of the top-surface Hf orbitals in the spin-up and spin-down channels, respectively. The grey background denotes the nonmagnetic bulk states. (b) Calculated 2D anomalous Hall conductivity (AHC) on the single surface of a 10-layer Hf$_2$S slab.}
\label{fig2}
\end{figure}

In summary, we have studied the electride Hf$_2$S in the monolayer and multilayer forms by using first-principles electronic structure calculations and Kubo formula approach. Our calculations indicate that in monolayer Hf$_2$S both the Hf atoms and the surface electron gases are spin polarized, which are ferromagnetic in the same Hf layer but antiferromagnetic between the upper and bottom Hf layers. After stacking to the multilayer, the interstitial electron gases between the neighboring Hf$_2$S layers become nonmagnetic, while the surface electron gases retain the original spin polarizations as in the monolayer case. The coexistence of nonmagnetic bulk states and ferromagnetic surface states may produce some interesting physics properties, such as the surface-related LHE and the non-zero surface AHC. The electride Hf$_2$S, as well as its isostructural compounds Hf$_2$Se~\cite{Zhang21} and Hf$_2$Te, can thus hold exotic quantum phenomena that are worth of further theoretical and experimental investigations.

We thank Xiao Zhang for helpful communication. This work was supported by the National Natural Science Foundation of China (Grants No. 12174443 and No. 11934020), the National Key R\&D Program of China (Grants No. 2017YFA0302903 and No. 2019YFA0308603), and the Beijing Natural Science Foundation (Grant No. Z200005). Computational resources were provided by the Physical Laboratory of High Performance Computing at Renmin University of China.

\textit{Note added}: In preparing our manuscript, we learned that a related work posted on the arXiv~\cite{ChoJH}, which also found the magnetic surface electron gas on Hf$_2$S but paid more attention to the topological properties.


\begin{thebibliography}{}
\bibitem{review1} H. Hosono and M. Kitano, Chem, Rev. {\bf 121}, 3121 (2021).
\bibitem{review2} X.-H. Yang and G.-C. Zhang, J. Phys. Chem. Lett. {\bf 11}, 3841 (2020).
\bibitem{concen} S. W. Kim, S. Matsuishi, T. Nomura, Y. Kubota, M. Takata, K. Hayashi, T. Kamiya, M. Hirano, H. Hosono, Nano Lett. {\bf 7}, 1138 (2007).
\bibitem{workfunc} Y. Toda, H. Yanagi, E. Ikenaga, J. J. Kim, M. Kobata, S. Ueda, T. Kamiya, M. Hirano, K. Kobayashi, H. Hosono, Adv. Mater. {\bf 19}, 3564 (2007).
\bibitem{emission} R. H. Huang, J. L. Dye, Chem. Phys. Lett. {\bf 166}, 133 (1990).
\bibitem{catalysts} J. Wu, J. Li, Y. Gong, M. Kitano, T. Inoshita, H. Hosono, Angew. Chem., Int. Ed. {\bf 58}, 825 (2019).
\bibitem{emitter} E. Feizi, A. K. Ray, J. Disp. Technol. {\bf 12}, 451 (2016).
\bibitem{supercon} M. Miyakawa, S. W. Kim, M. Hirano, Y. Kohama, H. Kawaji, T. Atake, H. Ikegami, K. Kono, and H. Hosono. J. Am. Chem. Soc. {\bf 129}, 7270 (2007).
\bibitem{Gd2C1} S. Y. Lee, J.-Y. Hwang, J. Park, C. N. Nandadasa, Y. Kim, J. Bang, K. Lee, K. H. Lee, Y.-W. Zhang, Y.-M. Ma, H. Hosono, Y. H. Lee, S.-G. Kim, and S. W. Kim, Nat. Commun. {\bf 11}, 1526 (2020).
\bibitem{wengsb} J.-C. Gao, Y.-T. Qian, H.-X. Jia, Z.-P. Guo, Z. Fang, M. Liu, H.-M. Weng, Z.-J. Wang, Sci. Bull. {\bf 67}, 598 (2022).
\bibitem{maprx} Y.-W. Zhang, H. Wang, Y.-C. Wang, L.-J. Zhang, and Y.-M. Ma, Phys. Rev. X {\bf 7}, 011017 (2017).
\bibitem{Y2C} H. Tamatsukuri, Y. Murakami, Y. Kuramoto, H. Sagayama, and H. Hosono, Phys. Rev. B. {\bf 102}, 224406 (2020).
\bibitem{topoprx} M. Hirayama, S. Matsuishi, H. Hosono, and S. Murakami, Phys. Rev. X {\bf 8}, 031067 (2018).
\bibitem{Ca2N1} D. L. Druffel, K. L. Kuntz, A. H. Woomer, F. M. Alcorn, J. Hu, C. L. Donley, and S. C. Warren, J. Am. Chem. Soc. {\bf 138}, 16089 (2016).
\bibitem{Ca2N} X.-L. Qiu, J.-F. Zhang, Z.-Y. Lu, and K. Liu, J. Phys. Chem. C {\bf 123}, 24698 (2019).

\bibitem{Gd2C} S. Liu, C. Wang, L. Liu, J.-H. Choi, H.-J. Kim, Y. Jia, C. H. Park, and J.-H. Cho, Phys. Rev. Lett. {\bf 125}, 187203 (2020).
\bibitem{Hf2S} S. H. Kang, J. Bang, K. Chung, C. N. Nandadasa, G. Han, S. Lee, K. H. Lee, K. Lee, Y. Ma, S. H. Oh, S.-G. Kim, Y.-M. Kim, S. W. Kim, Sci. Adv. {\bf 6}, eaba7416 (2020).

\bibitem{dft1} P. Hohenberg and W. Kohn, Phys. Rev. {\bf 136}, B864 (1964).
\bibitem{dft2} W. Kohn and L. J. Sham, Phys. Rev. {\bf 140}, A1133 (1965).
\bibitem{pwscf} P. Giannozzi, S. Baroni, N. Bonini, M. Calandra, R. Car, C. Cavazzoni, D. Ceresoli, G. L. Chiarotti, M. Cococcioni, I. Dabo, \textit{et al.}, J. Phys.: Condens. Matter {\bf 21}, 395502 (2009).
\bibitem{vasp1} G. Kresse and J. Furthm$\ddot{\text{u}}$ller, Comput. Mater. Sci {\bf 6}, 15 (1996).
\bibitem{vasp2} G. Kresse and J. Furthm$\ddot{\text{u}}$ller, Phys. Rev. B {\bf 54}, 11169 (1996).
\bibitem{pbe} J. P. Perdew, K. Burke, and M. Ernzerhof, Phys. Rev. Lett. {\bf 77}, 3865 (1996).
\bibitem{AHC} D. Xiao, M.-C. Chang, and Q. Niu, Rev. Mod. Phys. {\bf 82}, 1959 (2010).
\bibitem{wt} Q. Wu, S. Zhang, H.-F. Song, M. Troyer, and A. A. Soluyanov, Comput. Phys. Commun. {\bf 224}, 405 (2018).
\bibitem{mlwf} A. A. Mostofi , J. R. Yates, G. Pizzi, Y.-S. Lee, I. Souza, D. Vanderbilt, and N. Marzari, Comput. Phys. Commun. {\bf 185}, 2309 (2014).
\bibitem{sm} See Supplemental Material for more calculation details.
\bibitem{ELF} B. Silvi and A. Savin, Nature {\bf 371}, 683-686 (1994).
\bibitem{MBE} Z. Sun, X. Han, Z. Cai, S. Yue, D. Geng, D. Rong, L. Zhao, Y.-Q. Zhang, P. Cheng, L. Chen, X. Zhou, Y. Huang, K. Wu, B. Feng, Sci. Bull. in proof.
\bibitem{Zhang21} X. Wang, X. Qiu, C. Sun, X. Cao, Y. Yuan, K. Liu, and X. Zhang, Chin. Phys. Lett. {\bf 38}, 017302 (2021).
\bibitem{LHE} A. Gao, Y.-F. Liu, C. Hu, J.-X. Qiu, C. Tzschaschel, B. Ghosh, S.-C. Ho, D. BšŠrubšŠ, R. Chen, H. Sun, Z. Zhang, X.-Y. Zhang, Y.-X. Wang, N. Wang, Z. Huang, C. Felser, A. Agarwal, T. Ding, H.-J. Tien, A. Akey, J. Gardener, B. Singh, K. Watanabe, T. Taniguchi, K. S. Burch, D. C. Bell, B. B. Zhou, W. Gao, H.-Z. Lu, A. Bansil, H. Lin, T.-R. Chang, L. Fu, Q. Ma, N. Ni, and S.-Y. Xu, Nature {\bf 595}, 521 (2021).
\bibitem{FGTexp} Y. Deng, Y. Yu, Y. Song, J. Zhang, N. Z. Wang, Z. Sun, Y. Yi, Yi Z. Wu, S. Wu, J. Zhu, J. Wang, X. H. Chen, and Y. Zhang, Nature {\bf 563}, 94 (2018).
\bibitem{FGT} X. Lin and J. Ni, Phys. Rev. B {\bf 100}, 085403 (2019).
\bibitem{ChoJH} S. Liu, C. Wang, H. Jeon, Y. Jia, and J.-H. Cho, arXiv:2206.03689 (2022).


\end{thebibliography}
\end{document}